\documentclass[aps,prb,twocolumn]{revtex4-1}

\usepackage{graphicx}
\usepackage{dcolumn}
\usepackage{bm}
\usepackage{color}
\usepackage{amsmath}
\usepackage{amsfonts}
\usepackage{amssymb}
\usepackage{subfigure}
\usepackage{color}

\begin{document}

\title{Microscopic and continuum descriptions of Janus motor fluid flow fields}

\author{Shang Yik Reigh}
 \email{syr20@cam.ac.uk}
 \affiliation{Department of Applied Mathematics and Theoretical Physics, Centre for Mathematical Sciences,
  University of Cambridge, Wilberforce Road, Cambridge CB3 0WA, United Kingdom}
 \author{Mu-Jie Huang}
 \email{mjhuang@chem.utoronto.ca}
 \author{Jeremy Schofield}
 \email{jmschofi@chem.utoronto.ca}
 \author{Raymond Kapral}
 \email{rkapral@chem.utoronto.ca}
 \affiliation{Chemical Physics Theory Group, Department of Chemistry, University of Toronto, Toronto, Ontario M5S 3H6, Canada}





\begin{abstract}
Active media, whose constituents are able to move autonomously, display novel features that differ from those of equilibrium systems. In addition to naturally-occurring active systems such as populations of swimming bacteria, active systems of synthetic self-propelled nanomotors have been developed.  These synthetic systems are interesting because of their potential applications in a variety of fields. Janus particles, synthetic motors of spherical geometry with one  hemisphere that catalyzes the conversion of fuel to product and one noncatalytic hemisphere, can propel themselves in solution by self-diffusiophoresis. In this mechanism the concentration gradient generated by the asymmetric catalytic activity leads to a force on the motor that induces fluid flows in the surrounding medium. These fluid flows are studied in detail through microscopic simulations of Janus motor motion and continuum theory. It is shown that continuum theory is able to capture many but not all features of the dynamics of the Janus motor and the velocity fields of the fluid.
\end{abstract}






\maketitle

\section{Introduction}

Synthetic motors that transduce chemical energy into directed motion may be used as small-scale cargo delivery vehicles for applications in science and medicine, and the realization of novel nanoscale applications has stimulated much research activity in this area of science.~\cite{sen:04,ozin:05,wangbook:13,wang:13,guix:14,kapral:13,sanchez:14,colberg:14,sen:15,Ma_etal_15} Of the many different motors that have been constructed in the laboratory, our interest is in those that operate by self-diffusiophoresis where concentration gradients of reactant and products produced by localized catalytic activity on the motor surface give rise to forces that lead to directed motion of the motor. In particular, we consider spherical Janus motors with catalytic and noncatalytic faces, since they have perhaps the simplest geometry and have been studied often.~\cite{golestanian-1:07,showalter:10,palacci:10,ebbens:12,bocquet:12,wang:13,lee:14,poon:14,Ma_etal_15} Investigations of the dynamics of these motors, with sizes varying from nanometers to microns, are theoretically challenging since they operate out of equilibrium and experience strong thermal fluctuations.

Continuum theories are usually employed to describe the mechanism that is responsible for diffusiophoretic motor motion.~\cite{anderson:89,goles:05} In such macroscopic theories, the fluid in which the motor moves is modeled by the Navier-Stokes equations and the concentration fields of the solute species by reaction-diffusion equations. Interactions of the fluid with the motors are taken into account by boundary conditions on the motor surface. Very small nanomotors lie in the domain where the validity of a continuum description of the fluid dynamics and the influence of fluid motion on motor dynamics should be questioned and tested. Evidence for the breakdown of continuum theory can be found in molecular dynamics simulations of sphere-dimer motors~\cite{kapral07} with lengths of a few nanometers: fluid structure and dynamics is microscopically complex in the vicinity of the motor, although the fluid velocity fields in the motor vicinity exhibit structured flow patterns after averaging over fluctuations~\cite{colberg_epl:14}, similar to those predicted by continuum theories~\cite{reigh:15}.

In this paper we present results of coarse-grain microscopic simulations of Janus motors propelled by a self-diffusiophoretic mechanism. The Janus motors interact with the chemical species in the solvent through hard collisions~\cite{huang:16} and the solvent molecules evolve through multiparticle collision dynamics~\cite{male:99}. A particular focus of this study is the nature of the solvent flow fields near to and far from the Janus motor, since they are an integral part of the propulsion mechanism. The continuum equations, subject to boundary conditions on the motor, are solved to obtain the motor velocity and fluid velocity fields, so that the continuum and microscopic models may be compared.

\section{Microscopic and continuum descriptions}

\subsection{Microscopic dynamics}

In the microscopic simulations the properties of the entire system, comprising the Janus motor and the surrounding multi-component fluid medium, follow directly from the specification of the reactive and nonreactive interactions and numerical solution of the equations of motion. The coarse-grain microscopic dynamical scheme for the simulation of diffusiophoretic Janus motors studied here has been described in detail previously~\cite{debuyl:13,huang:16}, and only a brief outline will be given below.

\begin{figure}[htbp]
  \includegraphics[width=2.0in]{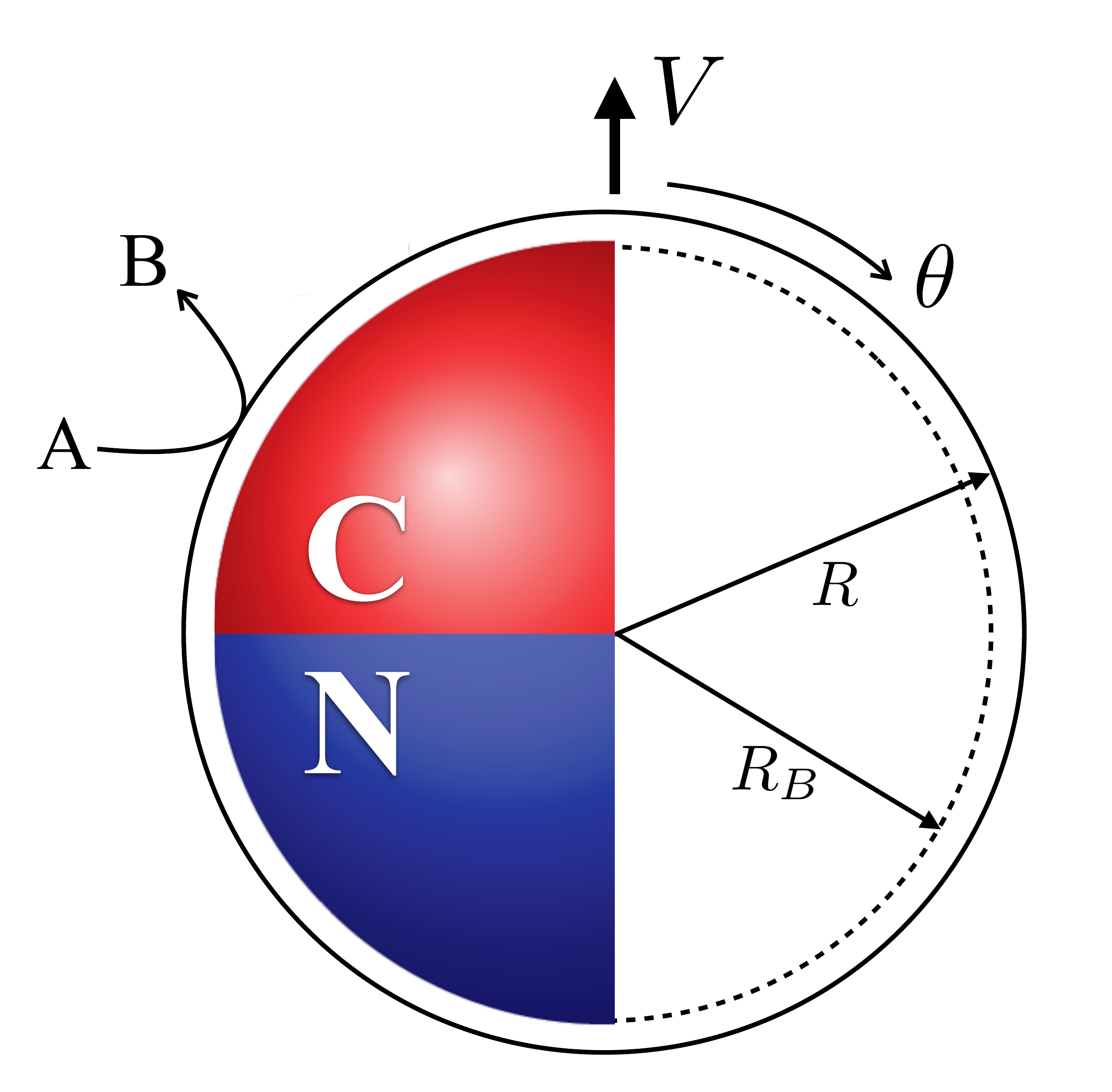}
  \caption{(Online version in colour)
  A Janus motor with catalytic ($C$) and non-catalytic ($N$) faces. The catalytic face converts fuel $A$ to product $B$. The motor is propelled in the direction of $\bm{V}=\hat{\bm{z}}V_z$ which is taken to lie along the polar $z$ axis in a spherical polar coordinate system with polar angle $\theta$. The left part of the figure shows the geometry used in the continuum model. The solid outer circle denotes the outer edge of the boundary layer at radius $R$. The right part of the figure shows the geometry used in the microscopic hard bounce-back model. The inner dashed circle indicates the radius at which bounce-back collisions of $B$ particles take place, while the outer solid circle, which coincides with the outer edge of the boundary layer, is the bounce-back radius for $A$ particles.
  }
  \label{fig_jan}
\end{figure}

The simulations are performed by molecular dynamics for a Janus motor, which is combined with multiparticle collision dynamics for fluid motion.~\cite{male:99,Malevanets_Kapral_00,kapral:08,gompper:09} In multiparticle collision dynamics the fluid is modelled by $N_s$ point particles of species $A$ and $B$ with mass $m$ in a cubic volume $L^3$ with linear size $L$. The fluid particles interact with the Janus motor through central intermolecular potentials, $W_\alpha(r)$, where $\alpha=A,B$ indicates the particle type interacting with the Janus motor. Interactions with the catalytic face of the Janus motor may lead to reactions that convert fuel molecules to products. Collision rules that specify the nature of the reaction and conserve mass, momentum and energy may be constructed.~\cite{debuyl:13} Here we suppose that the simple irreversible chemical reaction $A \rightarrow B$ takes place whenever an $A$ molecule collides with the catalytic hemisphere and exits the reaction surface with radius $R$. In this study we adopt a hard collision model for the reactive and nonreactive interactions with the Janus motor.~\cite{huang:16} The $A$ and $B$ species interact with the motor through hard potentials, $W_{\alpha}(r)=\infty$, for $r < R_{\alpha}$ and $W_{\alpha}(r)=0$ for $r \geq R_{\alpha}$, where $R_\alpha$ denotes the collision radius for species $\alpha$. We denote the larger of the radii by $R$ and assume that the collision radii are chosen so that $R -{R}_{\alpha}$ is small compared to the motor radius (see Fig.~\ref{fig_jan}). The chemical species $\alpha$ experiences modified bounce-back collisions at radius $R_\alpha$. In the remainder of the paper we take $R=R_A > R_B$ so that the Janus motor is propelled in a direction with the catalytic hemisphere at its head.

There are no intermolecular potentials among the solvent particles; their interactions are described by multiparticle collisions. The evolution of the system consists of streaming and collision steps. In the streaming steps the system evolves by Newton's equation of motion with impulsive forces determined from the intermolecular potentials of the solvent particles with the Janus motor. In the collision steps, which occur at time intervals $\tau$, the system is partitioned into cells $\xi$ of size $a$ which are assigned rotation operators, chosen from a set of rotation matrices. Particle velocities in a cell are then rotated relative to the center of mass velocity in a cell in order to obtain the post-collision velocity.
In order to maintain the system out of equilibrium fuel must be supplied and product removed. This may be done by explicit fluxes of these species at the boundaries of the system, or internally by out-of-equilibrium bulk reactions. Here we utilize bulk reactions to establish the nonequilibrium steady state for Janus motor operation. This is easily accomplished by using reactive multiparticle collision dynamics~\cite{rohlf:08}. Once the system is partitioned into cells for the multiparticle collisions, cell-level reactions that convert product $B$ particles back to fuel $A$ particles are carried out as follows:  at each collision step, the reaction, $B \stackrel{k_2}\rightarrow A$, takes place independently in each cell $\xi$ with probability $p^{\xi}(N_B^{\xi}) = 1 - e^{-a_2^{\xi}\tau}$, where $N_B^{\xi}$ is the total number of $B$ particles in cell $\xi$ and $a_2^{\xi} = k_2N_B^{\xi}$.

The important feature of this coarse-grain microscopic dynamics is that it conserves mass, momentum and energy. Consequently, the Navier-Stokes and reaction-diffusion equations can be derived from it, and it is able to describe the fluid flow fields and species concentration fields that accompany the motion of Janus motors. In addition analytical expressions for the transport properties of the solvent have been obtained;  in particular the solvent viscosity is known and this property is central to the motor propulsion and fluid flows that arise from motor motion.

\subsection{Continuum model}

In the continuum description we consider a macroscopically large spherical Janus motor with radius $R_J$ moving with velocity $\bm{V}$ in an incompressible  fluid with viscosity $\eta$. We further assume that the Reynolds number is small so that inertia may be neglected and the Stokes equations describe the fluid velocity fields. The concentration fields of the $A$ and $B$ species are described by reaction-diffusion equations with common species diffusion coefficient $D$, and the P\'eclet number is taken to be small so that advective effects may be neglected. Since the system is maintained in a steady state by nonequilibrium bulk reactions as described above for microscopic dynamics, these reactions are accounted for by mass action kinetic terms in the reaction-diffusion equations.

The computation of the velocity of motors propelled by the diffusiophoretic mechanism using continuum theory is well established~\cite{anderson:89,golestanian:07,julicher:09,popescu:10,sabass:11}, and the flow fields that result from a multipole description of swimming bodies have been studied~\cite{lauga:12}. The fluid velocity field is assumed to satisfy stick boundary conditions on the surface of the Janus particle. The potential functions $W_\alpha$ through which the fuel and product species interact with the Janus particle have a finite range. These interactions, in conjunction with the inhomogeneous concentration fields of species generated in the vicinity of the motor as a result of the asymmetric catalytic activity, give rise to body forces acting on the motor. Since momentum is conserved these forces generate fluid flows within the boundary layer of thickness $\delta$ where forces act, and the velocity field, $\bm{v}(R,\theta) \equiv  \bm{v}_s(\theta)$, at the outer edge of the boundary layer, $R_J+\delta \equiv R$, is the slip velocity, which takes the form~\cite{anderson:89},
\begin{equation} \label{slip}
  \bm{v}_s(\theta) = -\frac{k_BT}{\eta} \Lambda \bm{\nabla}_{\theta} c_B(R,\theta),
 \end{equation}
where $\Lambda = \int_0^\infty r(e^{-W_{B}(r)/k_BT}-e^{-W_{A}(r)/k_BT}) dr$. In writing this equation we used that fact that the total number of $A$ and $B$ particles is conserved in reactions so that the concentrations satisfy $c_0=c_A +c_B$, where $c_0$ is the total concentration of $A$ and $B$.

Making use of the reciprocal theorem of hydrodynamics~\cite{happel:73}, the velocity of the Janus motor can be obtained from the surface average of the slip velocity at the outer edge of the boundary layer at $R$: $\bm{V}= -\langle \bm{v}_s \rangle$. In order to compute the velocity, the solute concentration field at the outer edge of the boundary layer is needed. To find this field for our system with bulk reactions the reaction-diffusion equation, $ D \nabla^2 c_B -k_2 c_B = 0$, must be solved subject to the boundary condition,  $k_D R \frac{\partial c_A(r,\theta)}{\partial r}\Big|_{r=R}  =k_0 c_A(R,\theta) H(\theta)$. Here $k_0$ and $k_D = 4\pi RD$ are the intrinsic and Smoluchowski rate constants, and $H(\theta)$ is a characteristic function that is unity on the catalytic hemisphere ($0\leq\theta\leq\pi/2$) and zero on the noncatalytic hemisphere ($\pi/2\leq\theta\leq\pi$). The solution may be written as the series,
\begin{equation} \label{diff_sol}
  c_B = c_0\sum_{n=0}^{\infty}a_n f_n(r) P_n (\mu),
 \end{equation}
where $P_n(\mu)$ is a Legendre polynomial, $\mu=\cos \theta$ and $f_n(r)$ is determined from the solution of the radial equation. The $f_n(r)$ functions are defined such that $f_n(R)=1$. The unknown coefficients $a_n$ may then be determined from the solution of a linear set of algebraic equations~\cite{huang:16}. Substitution of Eq.~(\ref{diff_sol}) into the expression for the slip velocity and computation of the surface average yields the velocity component of the Janus motor along the director $\hat{\bm{z}}$ of the Janus motor, $V_z= \frac{2}{3} \frac{k_BT}{\eta R} \Lambda c_0 a_1$, and for the hard collision model $\Lambda$ is given by $\Lambda= (R_A^2-R_B^2)/2$. Comparisons of the measured motor velocity from simulations with this theoretical expression have been given earlier~\cite{huang:16}. We now turn to the main focus of this paper: the fluid velocity fields that accompany motor motion.

The velocity fields are an integral part of the propulsion mechanism. In addition to the information they provide about dynamical processes arising from the structure of the boundary layer, they are responsible for the hydrodynamic interactions among motors. The slip velocity is an important ingredient in the calculation of the fluid flow fields since it enters in  the boundary conditions that the flow fields must satisfy. In order to find the fluid velocity fields, we solve the Stokes equation,
$\nabla p = {\eta} \nabla^2 \bm{v}$, where $p$ is the pressure field, for an incompressible fluid satisfying $\nabla \cdot \bm{v}=0$ subject to the boundary conditions, $\bm{v}(R,\theta) = \bm{V}+\bm{v}_s$ at the outer edge of the boundary layer. In addition, far from the Janus motor we require that $\lim_{r \to \infty} \bm{v}(\bm{r}) = 0$. The details of this calculation are given in the Appendix and the resulting expressions for the radial $v_r$ and angular $v_{\theta}$ parts of the velocity field are found to be
\begin{widetext}
\begin{eqnarray} \label{vf}
    &&v_{r}(r,\mu) =\frac{k_BT}{\eta} \Lambda c_0\left[\frac{2 a_1 }{3R}\Big(\frac{R}{r}\Big)^3P_1(\mu)
    +\sum_{n=2}^{\infty} \frac{n(n+1) a_n}{2R}\Big(\frac{R}{r}\Big)^{n}
    \Big\{\Big(\frac{R}{r}\Big)^{2}-1\Big\}
    P_n(\mu)\right] \\
    &&v_{\theta}(r,\mu) = \frac{k_BT}{\eta} \Lambda c_0 \left[\frac{a_1 }{3R}\Big(\frac{R}{r}\Big)^3  V_1(\mu)
    +\sum_{n=2}^{\infty} \frac{n(n+1)a_n}{4R}\Big(\frac{R}{r}\Big)^{n}
    \Big\{n\Big(\frac{R}{r}\Big)^{2}-(n-2)\Big\}
     V_n(\mu)\right]. \nonumber
\end{eqnarray}
\end{widetext}
Here $V_n=-\frac{2}{n(n+1)}P_n^1(\mu)$, where $P_n^1$ is the associated Legendre polynomial of order $1$. Note that if one drops terms with $n \ge 2$ the flow field at distances far from the Janus particle is given by
\begin{align}
  \bm{v} = \frac{1}{2}\Big( \frac{R}{r} \Big)^3(3 \hat{\bm{r}}\hat{\bm{r}}-\bm{I})\cdot \bm{V},
\end{align}

where $\bm{I}$ the unit dyadic. This fluid flow field exhibits a dipolar flow pattern, which is generated by the slip velocity
$\bm{v}_s =\frac{k_BT}{\eta R} \Lambda  c_0 a_1 \sin\theta\hat{\bm{\theta}}$; hence, the dipolar flow is a special case of a general flow produced by the motion
of a self-diffusiophoretic \cite{anderson:86,anderson:89,yang14} and a thermophoretic Janus motor~\cite{Yang_Ripoll:13}.

The stream function may be obtained from the fluid velocity field and is given by (see Appendix)
\begin{widetext}
\begin{equation}\label{eq:stream}
  \psi(r,\mu) =-V_z \frac{R}{r}R^2 I_2(\mu)+\frac{k_BT}{2 \eta R} \Lambda  c_0 \sum_{n=2}^{\infty}n(n+1)a_n \left(\frac{R}{r}\right)^n (r^2-R^2 )I_{n+1}(\mu),
\end{equation}
\end{widetext}

where $I_n(\mu) = (P_{n-2}-P_n)/(2n-1)$ is the Gegenbauer function of the first kind~\cite{happel:73}. These flow fields will be discussed in the following section.

\section{Comparison of theory and simulations}\label{sc:comp}

In this section we compare the results of microscopic simulations with the predictions of macroscopic continuum theory for the properties of self-diffusiophoretic Janus motors. Our main focus is on the fluid velocity fields but results for the concentration fields will also be presented since they enter into the calculation of the velocity field. As discussed earlier, once the interaction model for the system is specified in the microscopic simulation, all transport properties, such as the solvent viscosity, species diffusion coefficients, Janus motor velocity and rotational and translational diffusion coefficients, follow directly from averages of dynamical quantities determined from solutions of the equations of motion of the entire system. In contrast, the continuum model requires transport properties as input, such as the viscosity, species diffusion constants and reaction rate constants which enter in the boundary conditions for the velocity and concentration fields. These boundary conditions are applied at $R$, the outer edge of the boundary layer. Consequently, in order to compare the predictions from microscopic simulations with continuum theory, the macroscopic parameters needed for the continuum calculation must be determined from the microscopic simulations.

The reaction boundary condition contains the Smoluchowski and intrinsic reaction rate constants, $k_D$ and $k_0$, respectively. These rate constants may be found by monitoring the time evolution of the concentrations of fuel $A$ molecules when only irreversible chemical reactions at the Janus motor are taken into account with no bulk reactions ($k_2 = 0$)~\cite{tucci:04}. From the rate equation $d c_A(t)/dt=-k(t)c_J c_A(t)$, where $c_J = 1/L^3 $ is the Janus motor number density,
one may determine $k(t)$, the time-dependent rate coefficient. Its initial value, determined from the initial slope of the $c_A(t)$ decay, is $k(0)=k_0$, the intrinsic rate constant. Its asymptotic value is  $\lim_{t \to \infty}k(t)=k=k_0 k_D/(k_0+k_D)$, from which we may determine $k_D$. The intrinsic rate constant $k_0$ may be estimated from kinetic theory as the product of the collision cross section and the mean speed of the $A$ particles, which gives $k_0 = {R_A^2} \sqrt{2\pi k_BT/m}$.  The Smoluchowski rate constant may also be estimated since $k_D= 4 \pi D R$ and the diffusion coefficient is known. As noted above, the expression for the solvent viscosity is also known, and all of the ingredients needed for a comparison with the continuum theory are available.

The following system parameters have been chosen for the microscopic simulations: we use dimensionless units where mass is in units of $m$, lengths in units of $a$, and energies in units of $k_BT$. Time is then expressed in units of $t_0=({ma^2/k_BT})^{1/2}$. In these units $R = 2.5$, $c_0\equiv N_s/L^3=10$, and the Janus motor has mass $M = \frac{4}{3}\pi R^3 c_0 m \approx 655$ and moment of inertia $I  \approx 1636$. The cubic simulation box  has linear dimension $L=50$ with periodic boundaries. By choosing the rotational angle $\alpha=120^{\circ}$ and collision time step $\tau = 0.1$ for multiparticle collisions, the fluid viscosity is $\eta = 7.93$ and the common self diffusion constant for the $A$ and $B$ particles is $D = 0.061$. The molecular dynamics time step for bounce-back collisions is $0.01$. The bulk reaction rate is chosen as $k_2=0.01$. The rate constant $k_D$ is found to be $k_D=1.94$ from simulation, in agreement with the theoretical estimate of $1.92$. The intrinsic rate coefficient from simulation is $k_0 \approx 15.0$ while the simple kinetic theory estimate is $15.7$ for $R_A = 2.5$. The $\Lambda$ factor for $R_B=2.485$ is given by $\Lambda=0.037$. Using these parameters, simulations yield an average propulsion speed of $V_S=0.0043$ while continuum theory predicts a value of $V_T=0.0064$. This propulsion speed, along with other parameter values, ensures that the Reynolds number is small, $Re=c_0RV_S/\eta\sim 0.013 \ll 1$, which implies that viscous effects dominate over inertial effects.  For these simulation conditions, the P\'eclet number is also small, $Pe=V_SR/D \sim 0.17 \ll 1$,
so that fluid advection is negligible in the reaction-diffusion equations while the Schmidt number is large, $Sc=\eta/c_0D \sim 13 \gg 1$,
which implies that momentum transport dominates over mass transport, characteristic of a liquid solvent. Results are obtained from averages over more than $200$ realisations of the dynamics.

\begin{figure}[htbp]
  \includegraphics[width=0.9\columnwidth]{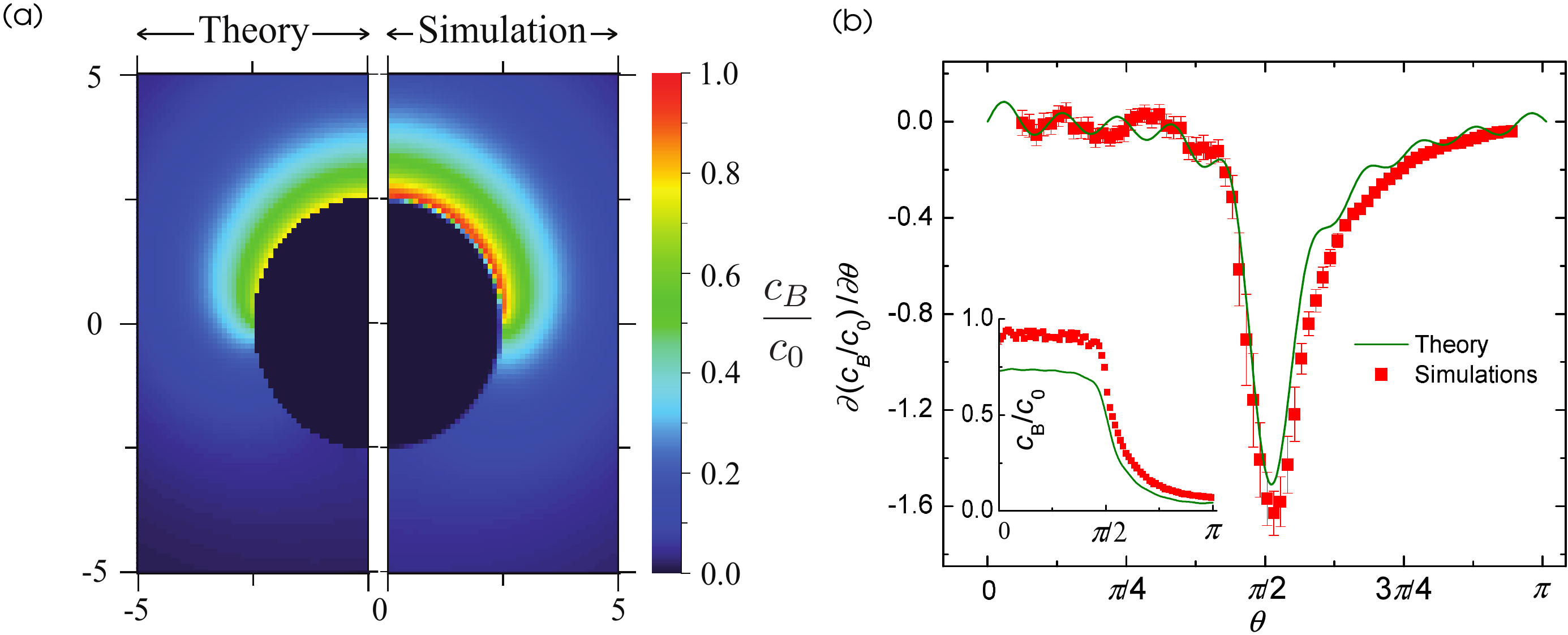}
  \caption{(Online version in colour)
    (a) Normalised concentration fields $c_B/c_0$ in the $xz$ plane ($y=0$);   continuum theory (left) and microscopic simulation (right).
    The upper hemisphere is the catalytic and the lower hemisphere is noncatalytic.
    (b) Quantitative comparison of concentration gradients near the edge of the boundary layer.
    The normalised concentration gradients $\partial (c_B/c_0)/\partial \theta$ are plotted as a function of the polar angle $\theta$.
    The solid green line is the continuum theory and the red squares are microscopic simulations.
    The inset shows the normalised concentration $c_B/c_0$ versus the angle $\theta$.
  }
  \label{fig_conc}
\end{figure}
First, we compare concentration fields in the vicinity of the Janus motor computed from continuum theory with those obtained from microscopic simulations. Figure~\ref{fig_conc} (a) shows the normalised concentration fields of the product $B$ particles ($c_B/c_0$) plotted in the $xz$ plane. Although there is overall qualitative agreement between the continuum and simulated concentration fields, the continuum model systematically underestimates the simulated values. 
This discrepancy could be due to the fact that the simulations are carried out at small but finite P\'{e}clet number, while the P\'{e}clet number in continuum model is assumed to be zero; however, simulations of a reactive Janus particle without propulsion ($R_A=R_B$) also show a similar discrepancy. Thus, the results point to a breakdown of the continuum model on small length scales near the surface of the particle where the catalytic reaction occurs. The continuum model accounts for reactions through a boundary condition on the outer edge of the boundary layer and further assumes the fluid flow field satisfies stick boundary conditions on the surface of the particle. Neither  of  these
conditions  is  fully  valid  when  one  examines  the  particle  dynamics  within  the  boundary  layer
in the microscopic simulations of our small motors. This breakdown of the continuum model is also reflected in the fact that the continuum and simulation values of the motor velocity differ. The microscopic model accounts for both of these effects through the direct hard intermolecular interactions between the fluid particles and the Janus motor. Figure~\ref{fig_conc} (b) presents quantitative comparisons of $B$ particle concentrations and concentration gradients near the boundary layer ($r \simeq R$) as a function of $\theta$. Although the concentration fields obtained from the continuum model and simulations present small differences as noted above, the concentration gradients responsible for the motor propulsion are in very good agreement.
\begin{figure}
\centering
  \includegraphics[width=1.0\columnwidth]{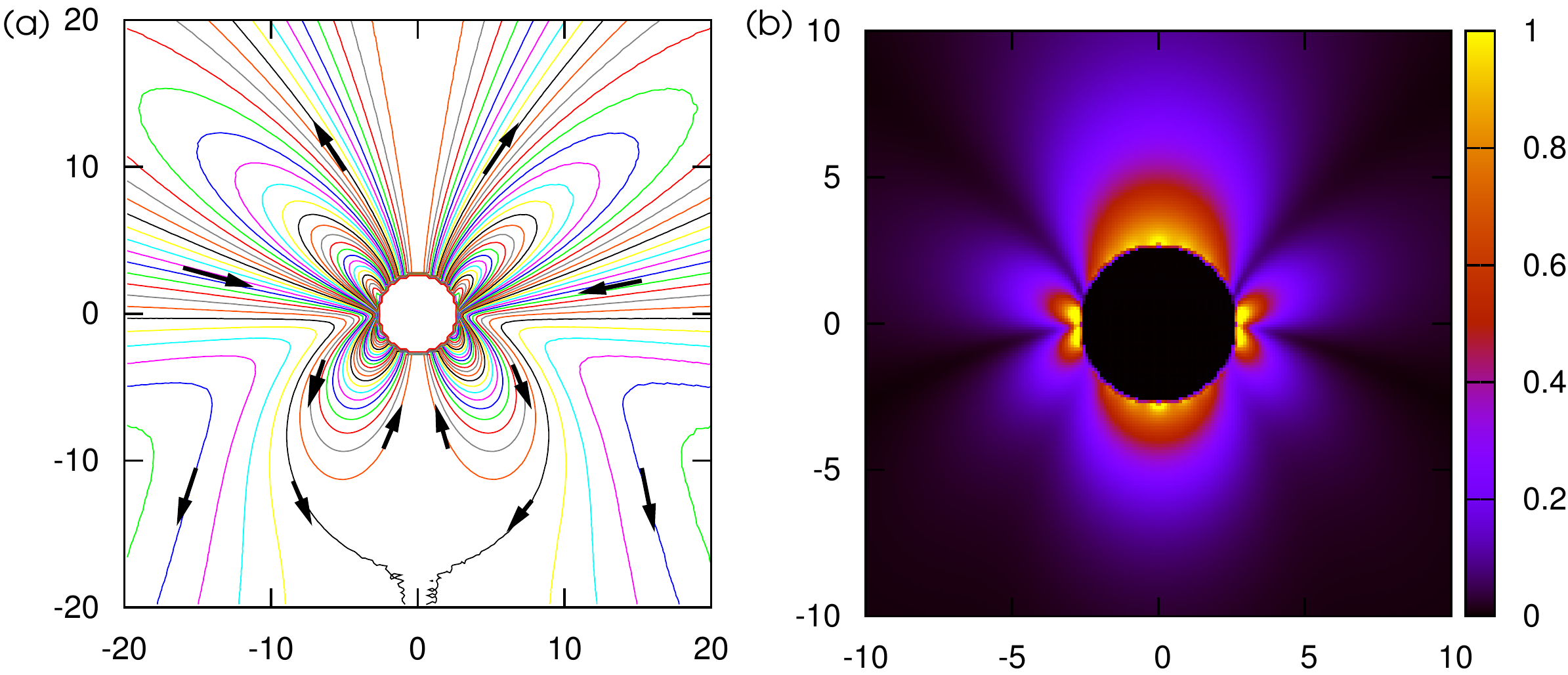}
  \caption{(Online version in colour)
  Streamlines and velocity field magnitude from continuum theory.    (a) Streamlines in the vicinity of the Janus motor in the laboratory frame where the motor moves with the velocity $\bm{V}$.  (b) Magnitude of the fluid velocity field scaled by propulsion velocity $V_T$ in the $xz$ plane.
  }
  \label{fig_stream}
\end{figure}

The motor moving with velocity $\bm{V}$ induces a flow field in the solvent. The characteristic patterns predicted from the Stokes equations can be seen in Fig.~\ref{fig_stream} (a). Note that the fluid flow field near the Janus motor does not have a dipolar form. Instead, it exhibits structure involving  more complicated fluid circulation.
Fluid is pushed from the front of the motor and returns in the lateral directions producing fluid circulation. Fluid entering from the side moves to the rear parts of the motor. Interesting fluid circulation patterns appear again near the rear of the motor. At large distances the fluid does not return to the motor but moves away. At mid-range distances where $r/a_0 \sim 19$, there is a stagnation point where the outgoing flow and incoming flow meet. On the basis of the far-field flow characteristics, the motor behaves as a pusher swimmer when $R_A>R_B$, since the fluid is pushed from the front and back and returns to the side of the motor~\cite{lauga:09}. When the molecular interactions change, e.g., $R_A<R_B$ so that $\Lambda<0$, the motor behaves as a puller swimmer, since fluid is pulled from the front and back and moves away on the sides.
Figure~\ref{fig_stream} (b) shows the magnitude of the fluid velocity scaled by the propulsion speed in the laboratory frame. There are strong outgoing and incoming flows on the front, sides and rear parts of the motor. Flow fields with similar characteristics have been reported for systems with motors propelled by thermophoresis~\cite{bickel:13,dmitry:15}.

\begin{figure}[htbp]
  \includegraphics[width=1.0\columnwidth,angle=0]{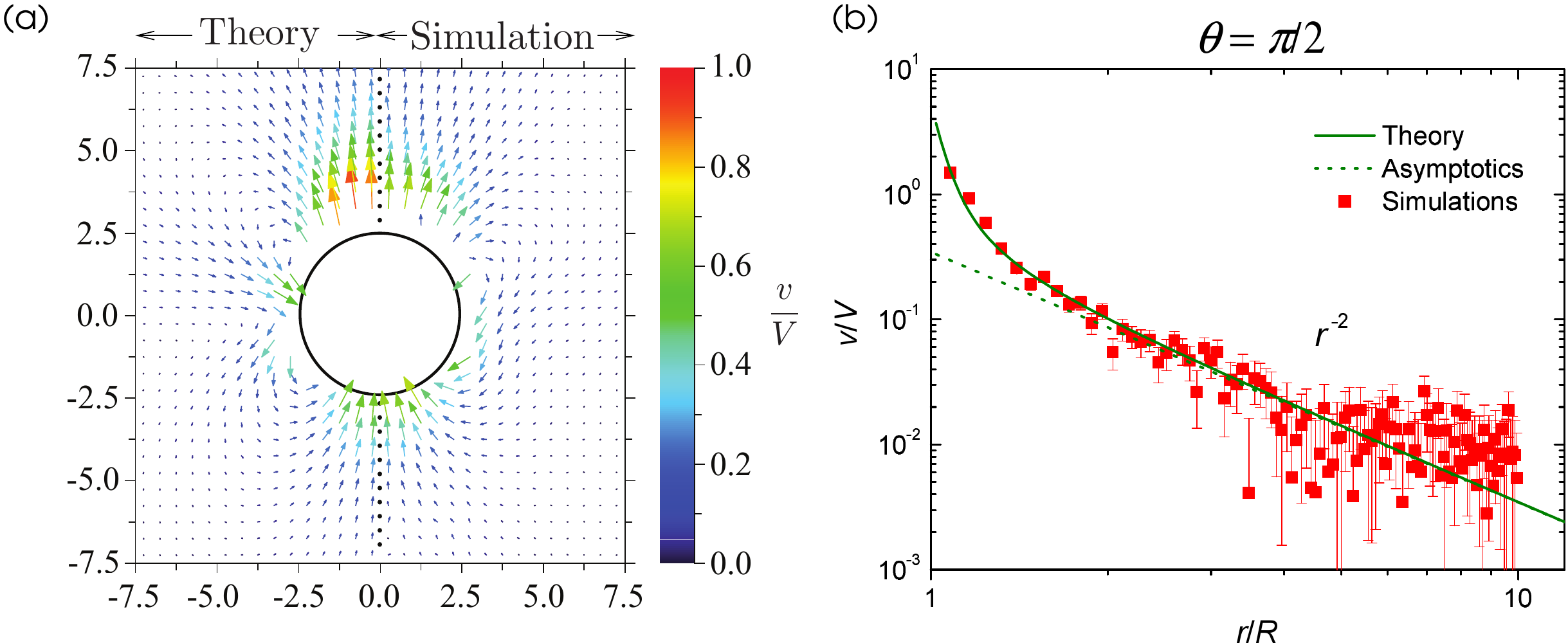}
  \caption{(a) Fluid velocity field: continuum theory (left) and microscopic simulations (right). The fluid velocity $v$ is scaled by the propulsion velocity $V$ ($V_T$ for the continuum theory and by $V_S$ for the microscopic simulations).
    (b) Magnitude of fluid velocity as a function of distance $r$ along the $\theta=\pi/2$ direction from the center of the motor.
    The solid green line is continuum theory, the dotted green line is the asymptotic value,  and the red squares are microscopic simulations.
    The $1/r^2$ power-law decay of velocity field is captured in the microscopic simulations. The radius of the motor is $R=2.5$.
  }
  \label{fig_vf}
\end{figure}
In Fig.~\ref{fig_vf} (a) the fluid velocity vector fields in the vicinity of the motor predicted by continuum theory and obtained from microscopic simulations are compared.
Overall the qualitative agreement between the velocity fields is good although the absolute magnitude of the velocity is somewhat different. The magnitude of the fluid velocity in the $\theta=\pi/2$ direction from the center of the motor, scaled by the magnitude of the motor velocity, is shown in Fig.~\ref{fig_vf} (b). The results of the continuum theory agree well with the simulation results in the near and far fields, although there are large fluctuations in the simulation results at large distance that arise from the effects of thermal noise on the small values of the velocity. Continuum theory predicts that the magnitude of the fluid velocity decays asymptotically as
$v = \sqrt{v_r^2+v_\theta^2} \sim \vert 3\frac{k_BT}{\eta R} \Lambda c_0 a_2 P_2(\mu) \vert (R/r)^2 + O(1/r^3)$ (see Eq.~\ref{vf}), and the $1/r^2$ power-law behavior is consistent with the far-field characteristics of pusher and puller swimmers as shown in the streamlines in Fig.~\ref{fig_stream}. The fluid velocity field obtained from the microscopic simulations confirms the $1/r^2$ power-law decay. Extensive averaging of instantaneous local velocities is required to obtain the structure of flow fields in the microscopic simulations. Since the velocity field is related to derivatives of the stream function, much more extensive simulations are required to obtain accurate values of the stream function suitable for comparison with the theoretical values.

\begin{figure}[htbp]
  \includegraphics[width=1.0\columnwidth,angle=0]{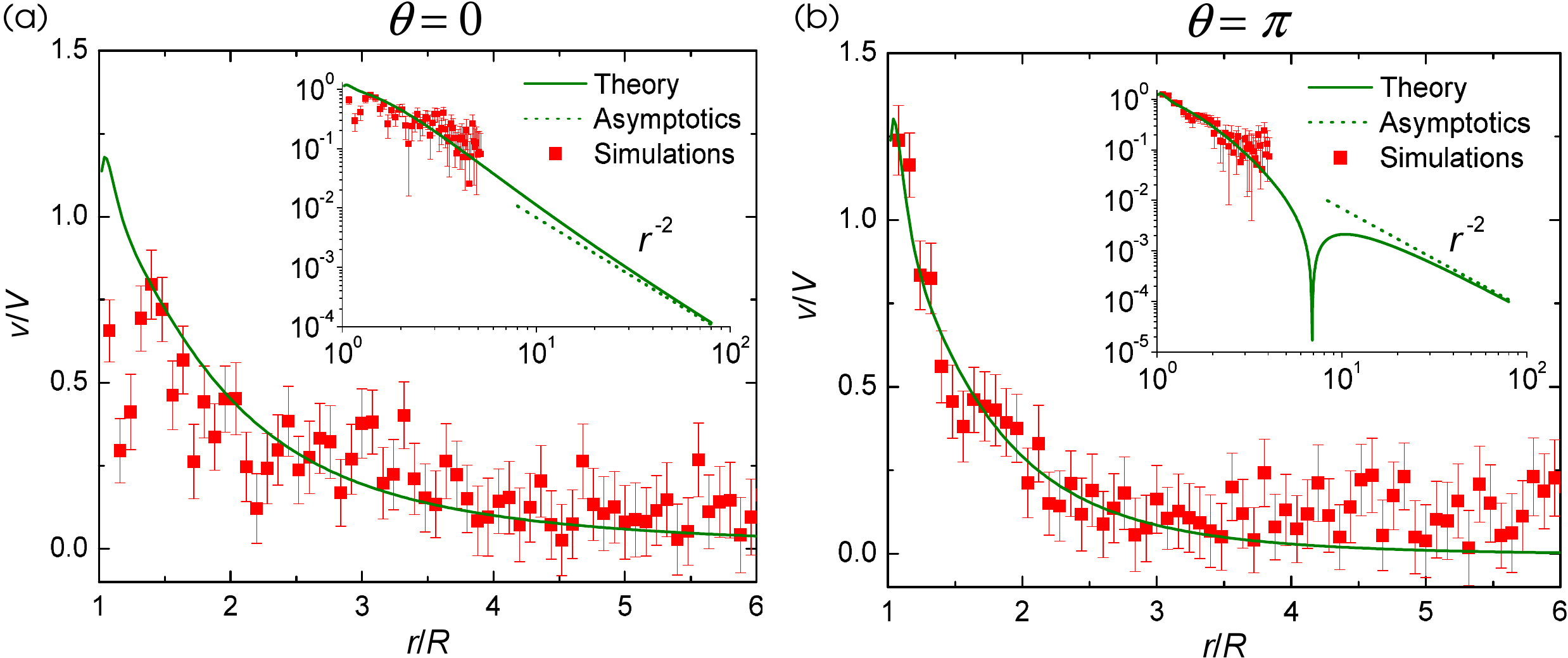}
  \caption{Comparison of continuum theory and simulations for the magnitude of the fluid velocity field. The fluid velocity fields $v$, scaled by the propulsion velocity $V$ ($V_T$ for theory and $V_S$ for simulations), are plotted as a function of distance $r$ along the (a) $\theta=0$ (b) $\theta=\pi$ directions from the center of the motor. Labeling is the same as that in Fig.~\ref{fig_vf}.
  }
  \label{fig_vf2}
\end{figure}
Figures~\ref{fig_vf2} (a) and (b) show quantitative comparisons of continuum theory and simulations for the magnitude of the fluid velocity field, scaled by the magnitude of the propulsion velocity, along the front ($\theta=0$) and back ($\theta=\pi$) directions from the center of the motor. Although the microscopic simulations exhibit large fluctuations, the continuum theory does capture major trends seen in the simulation results. Near the motor ($r/R \leq 1.25$), the discrepancies between the continuum theory and simulations are observed in the forward direction ($\theta=0$), while in the backward direction ($\theta=\pi$) the results are consistent. These differences can be attributed to approximations on the fluid flow fields within the boundary layer in the continuum model.

As shown in the insets of Fig.~\ref{fig_vf2}, continuum theory predicts a $1/r^2$ power-law decay at long distances. The distance at which this asymptotic regime is reached depends on the angle,
and analysis of the analytical formulas in Eqs.~(\ref{vf}) shows that this regime is reached approximately
one hundred times earlier for $\theta=\pi/2$ than for $\theta=0$ and $\theta=\pi$.
The velocity field obtained from the microscopic simulations is consistent with these observations. The results in Fig.~\ref{fig_vf} (b) for $\theta=\pi/2$ show that the simulation results lie in the asymptotic regime for $r/R>2$ while Fig.~\ref{fig_vf2} indicates that our simulations have not reached the asymptotic regime for the angles $\theta=0$ and $\theta=\pi$. Note also that in the inset of Fig.~\ref{fig_vf2} (b), one sees the stagnation point predicted from the streamlines in Fig.~\ref{fig_stream}, which occurs at distances that are inaccessible in the current simulation.

\section{Conclusion}

The microscopic simulations of the fluid velocity fields presented above show clear evidence of coherent hydrodynamic flows generated by Janus motor motion that are generally in accord with the predictions of continuum theory. Such agreement, while perhaps expected since these flows have their origin in the force-free nature of the dynamics and momentum conservation, is nevertheless interesting for several reasons. The small motors investigated in this study are subject to strong thermal fluctuations. The fluid flow fields presented in the comparisons were obtained by extensive averaging of local particle velocities subject to strong thermal fluctuations. Collective hydrodynamic effects in microscopic simulations have been observed much earlier in connection with long-time tails in velocity correlation functions, and our results show that such fields exist in far-from-equilibrium systems with active self-propelled particles. The continuum model utilizes a highly idealized treatment of the boundary layer and assumes no-slip boundary conditions at the motor surface and negligible fluid advection. Nevertheless, when suitable input parameters are used, notably the effective radius which characterizes the boundary layer, major features of the average flow fields are described well. The results in this paper should prove useful when more complex situations involving many motors are considered, since hydrodynamic interactions play a part in determining the dynamics and microscopic simulations provide a way to capture complex flow-field effects.

\section*{Appendix}

The fluid velocity fields in the region outside the boundary layer may be obtained by solving the Stokes equations. Taking the divergence of Stokes equations, $ \nabla p = {\eta} \nabla^2 \bm{v}$  and using the incompressibility condition, $\nabla \cdot \bm{v}=0$, the pressure field satisfies Laplace equation, $\nabla^2 p =0$. Since the pressure field is uniform far from the Janus motor, this axisymmetric field may be written in the form $p=p_\infty+\sum_{n=0}^{\infty} p_n$, where $p_\infty$ is the pressure far from the motor and $p_n= \gamma_n r^{-(n+1)} P_n(\mu)$ is a solid spherical harmonic. The general solution for an axisymmetric velocity field that vanishes far from the Janus motor may be written as~\cite{lamb}
\begin{align}
  \bm{v} = \sum_{n=1}^{\infty} \Big[\nabla \phi_n  +Z_n^{(1)}r^2\nabla p_n +Z_n^{(2)}\bm{r}p_n \Big],
  \label{gensol}
\end{align}
where $\phi_n=\chi_n r^{-(n+1)} P_n(\mu)$ is a solid spherical harmonic arising from the homogeneous equation
$\nabla^2 \bm{v}=0$. Substitution of these expressions into the Stokes equations, and using the incompressibility condition yields the following expressions for the $Z_n$ coefficients: $Z_n^{(1)}=(2-n)/(2\eta n(2n-1))$ and $Z_n^{(2)}= (n+1)/(\eta n(2n-1))$. The radial and angular components of the velocity field may then be written as
\begin{align}
  \begin{split}
    &v_r = \sum_{n=1}^{\infty}\Big[\frac{(n+1)}{2\eta(2n-1)}\frac{\gamma_n}{r^{n}}    -(n+1) \frac{\chi_n}{r^{n+2}}\Big] P_n(\mu),\\
    &v_{\theta} = \sum_{n=1}^{\infty} \Big[\frac{(n-2)}{2}\frac{(n+1)}{2\eta(2n-1)} \frac{\gamma_n}{r^{n}}
    - \frac{n(n+1)}{2}  \frac{\chi_n}{r^{n+2}} \Big]V_n(\mu).
    \label{lamb2}
  \end {split}
\end{align}

The coefficients $\chi_n$ and $\gamma_n$ may be determined from the boundary condition for the velocity fields at the outer edge of the boundary layer, $\bm{v}(R,\theta) = \bm{V}+\bm{v}_s(\theta)$, and the condition that the system is force-free. Using the expression for the concentration field given by Eq.~(\ref{diff_sol}), the slip velocity may be written as a series in $V_n(\mu)$: $\bm{v}_s(\theta)= \hat{\mbox{\boldmath{$\theta$}}}\frac{k_BT}{2 \eta R} \Lambda c_0 \sum_{n=1}^\infty n(n+1)a_n V_n(\mu)$, where $ \hat{\mbox{\boldmath{$\theta$}}}$ is a unit vector in the $\theta$ direction. The radial and angular components of the boundary condition take the forms,
\begin{widetext}
\begin{align}
  \begin{split}
    &v_r(R,\theta) = V_z P_1(\mu),\\
    &v_{\theta}(R,\theta) = \Big(\frac{k_BT}{\eta R} \Lambda c_0 a_1-V_z \Big) V_1(\mu)
    + \frac{k_BT}{2 \eta R} \Lambda c_0 \sum_{n= 2}^{\infty} n(n+1)  a_nV_n(\mu).
    \label{bc_sph}
  \end {split}
\end{align}
 \end{widetext}
Comparing Eqs.~(\ref{lamb2}) and (\ref{bc_sph}) for the $n=1$ terms yields the equations,
\begin{equation}
    - \frac{2 \chi_1}{R^3} + \frac{\gamma_1}{\eta R} = V_z,\quad
    -\frac{\chi_1}{R^3} - \frac{\gamma_1}{2\eta R} = \frac{k_BT}{\eta R} \Lambda c_0a_1 - V_z.
    \label{udc}
 \end{equation}

The total force ${\bf F}$ within the boundary layer is zero so that ${\bf F}=\int_S \bm{\Pi}\cdot \hat{\bm{r}}dS=0$, where the integral is over the boundary surface, the stress tensor is $ \bm{\Pi} = -p\bm{1} +\eta (\nabla {\bf v})^{sym}$, and the superscript denotes the symmetric gradient. Using the forms for the pressure and velocity fields and performing the surface integral, one finds $\gamma_1=0$. From Eq.~(\ref{udc}) it follows that $V_z=\frac{2}{3} \frac{k_BT}{\eta R } \Lambda  c_0a_1$, which agrees with the result obtained earlier in the text using the reciprocal theorem. Comparing coefficients for $n \ge 2$ terms one obtains,
\begin{align}
    \notag \gamma_n &= - k_BT \Lambda c_0 a_n  n (2n-1)   R^{n-1}  , \quad \\
   \chi_n &= -\frac{k_BT}{\eta}  \Lambda c_0 a_n  \frac{n}{2} R^{n+1}.
\end{align}
Substituting these results into the expression for the velocity field we find Eq.~(\ref{vf}) given in the main text.

The stream function may be obtained from the fluid velocity field. A general expression for the stream function for an unbounded fluid which does not diverge along the $z$ axis ($\mu=\pm 1$)
may be written as
\begin{align}
  \psi(r,\mu) = \sum_{n=1}^{\infty}\Big\{\frac{A_n}{r^{n-3}} + \frac{B_n}{r^{n-1}}  \Big\}I_n(\mu),
\end{align}
where $I_n(\mu) = (P_{n-2}-P_n)/(2n-1)$ is the Gegenbauer function of the first kind~\cite{happel:73}. The fluid velocity expressed in the spherical polar coordinate system is determined from the stream function by
\begin{align}
  \bm{v} = -\frac{1}{r^2\sin\theta}\frac{\partial \psi}{\partial \theta} \hat{\bm{r}}
  +\frac{1}{r\sin\theta}\frac{\partial \psi}{\partial r} \hat{\bm{\theta}}.
\end{align}
From the properties of the Gegenbauer functions~\cite{happel:73}, $\frac{\partial I_{n}}{\partial \mu} = -P_{n-1}$ and $(1-\mu^2)\frac{\partial P_{n-1}}{\partial \mu} = n(n-1)I_{n}$, the radial and tangential components of the velocity are given by
\begin{align}
     v_r &= \sum_{n=1}^{\infty}\Big[-\frac{A_{n+1}}{r^n}-\frac{B_{n+1}}{r^{n+2}}\Big]P_n(\mu),\quad \\
    v_{\theta} &= \sum_{n=1}^{\infty}\Big[-\frac{n-2}{2}\frac{A_{n+1}}{r^n} - \frac{n}{2}\frac{B_{n+1}}{r^{n+2}} \Big]V_n(\mu).
\end{align}
Comparing these expressions for the components of the velocity field with those in Eq.~(\ref{lamb2}),
one finds that the coefficients appearing in the stream function are related by $A_1=B_1=0$, $A_2=0$, $B_2=-R^3V_z$ and $A_{n+1} =-(n+1)\gamma_n/(2\eta (2n-1))$ and $B_{n+1} =(n+1) \chi_n$ for $n \ge 2$, leading to Eq.~(\ref{eq:stream}) of the main text.

\enlargethispage{20pt}





\section*{Acknowledgements}
This work was supported in part by grants from the Natural Science and Engineering Council of Canada.  Computations were performed on the GPC supercomputer at the SciNet HPC Consortium. SciNet is funded by: the Canada Foundation for Innovation under the auspices of Compute Canada; the Government of Ontario; the Ontario Research Fund - Research Excellence; and the University of Toronto.




%







\end{document}